\documentclass[aps,twocolumn,showpacs, showkeys,nofootinbib,floatfix]{revtex4-1}

\usepackage{amsmath}
\usepackage{amsfonts}
\usepackage{txfonts}
\usepackage{graphicx}
\usepackage{dcolumn}
\usepackage{bbm}
\usepackage{amssymb}
\usepackage{latexsym}
\usepackage{titlesec}
\usepackage[colorlinks=true, linkcolor=red, citecolor=blue]{hyperref}
\usepackage{longtable}

\begin{document}

\title{Detecting Primordial Gravitational Waves Signal from BICEP2 and {\it Planck} HFI $353$GHz Dust Polarization}
\author{Lixin Xu}
\email{Corresponding author: lxxu@dlut.edu.cn}

\affiliation{Institute of Theoretical Physics, School of Physics \&
Optoelectronic Technology, Dalian University of Technology, Dalian,
116024, P. R. China}

\affiliation{College of Advanced Science \& Technology, 
Dalian University of Technology, Dalian, 116024, P. R. China}

\affiliation{State Key Laboratory of Theoretical Physics, Institute of Theoretical Physics, Chinese Academy of Sciences, Beijing, 100190, P. R. China}

\begin{abstract}
The dust polarization  is parameterized as a power law form of the multipole $l$: $D^{XX}_{l}=A^{XX}l(l+1)l^{\alpha_{XX}}/(2\pi)$ ($XX$ denotes $BB$ or $EE$), where $A^{XX}$ is its amplitude with the ratio $A^{BB}/A^{EE}=0.52\pm 0.02$ and $\alpha_{BB,EE}=-2.42\pm 0.02$. Extrapolating to $150$GHz from $353$GHz yields a value of $D^{BB}_{l=80}=(1.32\pm 0.29)\times 10^{-2}\mu K^2$ (and an additional uncertainty $(+0.28,-0.24)\times 10^{-2}\mu K^2$) over the range $40<l<120$. Based on these data, we report the tensor-to-scalar ratio $r=A_{t}/A_{s}$ defined at $k_0=0.05 \text{Mpc} ^{-1}$ by joining the BICEP2+{\it Planck}2013+WMAP9+BAO+HST and {\it Planck} HFI $353$GHz dust polarization and its implication to the detection of the primordial gravitational waves. Considering the $\Lambda$CDM+$r$ model, we found $r<0.108$ at $95\%$ confidence level with $\sigma_{stat}=0.29$ and $r<0.129$ at $95\%$ confidence level with $\sigma_{stat+extr}=0.29+0.28$. The results imply no significant evidence for the primordial gravitational waves in $1\sigma$ regions. However the post probability distribution of $r$ peaks at a small positive value. And $r$ moves to larger positive values when the extrapolation error bars are included. This might imply a very weak signal of the primordial gravitational waves. It also implies the crucial fact in calibrating the amplitude of the dust polarizations in detecting the primordial gravitational waves in the future. When the running of the scalar spectral tilt is included, we found $r<0.079$ at $95\%$ confidence level with $\sigma_{stat}=0.29$ and $r=0.091_{-0.069}^{+0.042}$ at $95\%$ confidence level with $\sigma_{stat+extr}=0.29+0.28$. The later one implies the detection of the primordial gravitational waves in $1\sigma$ regions at the cost of decreasing the value of $D^{BB}_{l=80}$ to $0.67_{-0.25}^{+0.25}$.
\end{abstract}



\maketitle

\section{Introduction}

The Background Imaging of Cosmic Extragalactic Polarization (BICEP2) experiment \cite{ref:BICEP21,ref:BICEP22} has detected the B-modes of polarization in the cosmic microwave background, where the tensor-to-scalar ratio $r=0.20^{+0.07}_{-0.05}$ with $r=0$ disfavored at $7.0\sigma$ of the lensed-$\Lambda$CDM model was found. However, it was debated that what BICEP2 detected is not the signal of the primordial gravitational waves but is the contamination coming from the foreground \cite{ref:Mortonson,ref:Flauger,ref:Colley}. Recently, {\it Planck} group measured the dust angular power spectrum of $D^{XX}_{l}=l(l+1)C^{XX}_l/(2\pi)$ ($XX$ denotes EE or BB) over the multipole range $40<l<600$ at intermediate and high Galactic latitudes form $100$GHz to $353$GHz. Extrapolation of the {\it Planck} $353$GHz data to $150$GHz gives almost the same magnitude as that of BICEP2 signal \cite{ref:Planckdust} in the range $40<l<120$. Although the measured ratio between the amplitudes of B and E modes polarization power spectra is $C^{BB}_{l}/C^{EE}_{l}=0.53$ which is not consistent with current theoretical models. The spectral energy distribution (SED) of the dust $D^{XX}_{l}$ from $100$GHz to $353$GHz is accurately described by the modified dust emission law that allows the extrapolation to $150$GHz of BICEP2. Therefore, it deserves to reanalyze the BICEP2 data including the possible detected dust polarization contamination. Actually,  it was done in \cite{ref:Huangdust} where $r<0.083-0.087$ at $95\%$ confidence level in the base $\Lambda$CDM+$r$ model without and with the extrapolation error bars were given.  

In this paper, we will reanalyze the tensor-to-scalar ratio in the following data sets: 

(i) For the {\it Planck} $353$GHz polarization data, we use the dust power $D^{XX}_{l}=l(l+1)C^{XX}_l/(2\pi)$ which is parameterized in the power law form of multipole $l$
\begin{equation}
D^{XX}_{l}=A^{XX}l(l+1)l^{\alpha_{XX}}/(2\pi),
\end{equation} 
where $XX$ denotes the B-mode or E-mode polarization, $A^{XX}$ is its amplitude. In this work, we take the corresponding model parameters with the following Gaussian priors: $D^{BB}_{l=80}=(1.32\pm 0.29)\times 10^{-2}\mu K^2$ (and an additional uncertainty $(+0.28,-0.24)\times 10^{-2}\mu K^2$ from the extrapolation over the range $40<l<120$), $\alpha_{BB,EE}=-2.42\pm 0.02$ and $A^{BB}/A^{EE}=0.53\pm 0.02$ instead of their central values \cite{ref:Planckdust}. One should notice that the extrapolation error $\sigma_{extr}=0.28\times 10^{-2}\mu K^2$ instead of $(+0.28,-0.24)\times 10^{-2}\mu K^2$ is adopted in this work. And the E-mode polarization is also included in our data analysis. 

(ii) The released BICEP2 CMB B-mode data \cite{ref:BICEP21,ref:BICEP22}.

(iii) The full information of CMB which include the recently released {\it Planck} data sets which include the high-l TT likelihood ({\it CAMSpec}) up to a maximum multipole number of $l_{max}=2500$ from $l=50$, the low-l TT likelihood ({\it lowl}) up to $l=49$ and the low-l TE, EE, BB likelihood up to $l=32$ from WMAP9, the data sets are available on line \cite{ref:Planckdata}.

(iv) For the BAO data points as 'standard ruler', we use the measured ratio of $D_V/r_s$, where $r_s$ is the co-moving sound horizon scale at the recombination epoch, $D_V$ is the 'volume distance' which is defined as
\begin{equation}
D_V(z)=[(1+z)^2D^2_A(z)cz/H(z)]^{1/3},
\end{equation}
where $D_A$ is the angular diameter distance. The BAO data include $D_V(0.106) = 456\pm 27$ [Mpc] from 6dF Galaxy Redshift Survey \cite{ref:BAO6dF}; $D_V(0.35)/r_s = 8.88\pm 0.17$ from SDSS DR7 data \cite{ref:BAOsdssdr7}; $D_V(0.57)/r_s = 13.62\pm 0.22$ from BOSS DR9 data \cite{ref:sdssdr9}. Here the BAO measurements from WiggleZ are not included, as they come from the same galaxy sample as $P(k)$ measurement.

(v) The present Hubble parameter $H_0 = (73.8\pm 2.4)\text{km s}^{-1} \text{Mpc}^{-1}$ from HST \cite{ref:HST} is used.

The scalar and tensor mode power spectra are parameterized as
\begin{eqnarray}
P_{s}(k)&\equiv& A_s(k/k_0)^{n_s-1+\frac{1}{2}\alpha_s\ln(k/k_0)},\\
P_{t}(k)&\equiv& A_t(k/k_0)^{n_t},
\end{eqnarray} 
where $n_s-1$ and $n_t$ are tilts of power spectrum of scalar and tensor modes, $k_0=0.05 \text{Mpc} ^{-1}$ is the pivot scale, and $\alpha_s=d n_s/d\ln k$ is the running of the scalar spectral tilt. The primordial tensor-to-scalar ratio is defined by $r\equiv A_t/A_s$ at different pivot scale, here, they are $r$ defined at $k_0=0.05 \text{Mpc} ^{-1}$ and $r_{0.002}$ defined at $k_0=0.002 \text{Mpc} ^{-1}$. Adiabatic initial conditions and inflation consistence relation $n_t=-r/8$ were assumed in this paper. 

\section{Constrained Results} \label{sec:results}

\begin{center}
\begin{figure}[tbh]
\includegraphics[width=9cm]{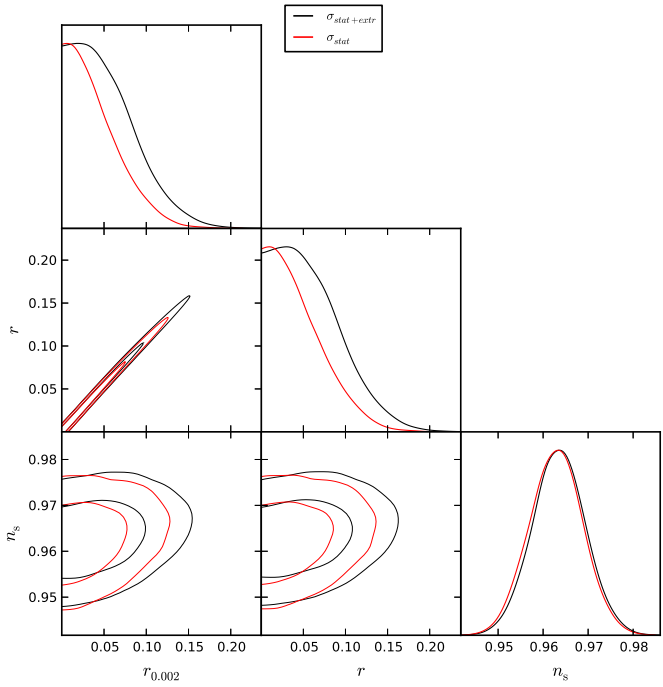}
\caption{The 1D marginalized distribution and 2D contours for $r_{0.002}$ and $n_s$ with $68\%$ C.L., $95\%$ C.L. by using {\it Planck}2013+{\it Planck}DUST+WMAP9+BICEP2+BAO+HST data points for $\Lambda$CDM+$r$ model with and without the extrapolation error bars.}\label{fig:contourI}
\end{figure}
\end{center}

We performed a global fitting to the model parameter space on the {\it Computing Cluster for Cosmos} by using the publicly available package {\bf CosmoMC} \cite{ref:MCMC}, which includes {\bf CAMB} \cite{ref:CAMB} to calculate the CMB power spectra. We modified the code to include the {\it Planck} dust polarization data and three new model parameters $D^{BB}_{l=80}$, $A^{BB}/A^{EE}$ and $\alpha_{XX}$ which priors are summarized in Table \ref{tab:results}. The running eight chains were stopped when the Gelman \& Rubin $R-1$ parameter $R-1 \sim 0.02$ was arrived; that guarantees the accurate confidence limits. The obtained results are summarized in Table \ref{tab:results} and Figure \ref{fig:contourI} for the contour plots with and without the extrapolation error bars.

\begin{widetext}
\begingroup                                                                                                                     
\begin{center}                                                                                                                  
\begin{table}[tbh]                                                                                                                   
\begin{tabular}{cccccc}                                                                                                            
\hline\hline                                                                                                                    
Parameters & Priors & Mean with errors ($\sigma_{stat}$) & Best fit  ($\sigma_{stat}$) & Mean with errors ($\sigma_{stat+extr}$) & Best fit  ($\sigma_{stat+extr}$) \\ \hline
$\Omega_b h^2$ & $[0.005,0.1]$  & $0.02208_{-0.00025}^{+0.00025}$ & $0.02194$ & $0.02209_{-0.00025}^{+0.00025}$ & $0.02207$\\
$\Omega_c h^2$ & $[0.01,0.99]$ & $0.1179_{-0.0017}^{+0.0016}$ & $0.1171$ & $0.1178_{-0.0017}^{+0.0017}$ & $0.1170$\\
$100\theta_{MC}$ & $[0.5,10]$ & $1.04153_{-0.00055}^{+0.00056}$ & $1.04166$ & $1.04151_{-0.00056}^{+0.00056}$ & $1.04193$\\
$\tau$ & $[0.01,0.8]$ & $0.067_{-0.018}^{+0.022}$ & $0.065$ & $0.069_{-0.018}^{+0.021}$ & $0.085$\\
${\rm{ln}}(10^{10} A_s)$ & $[2.7,4.0]$ & $3.038_{-0.036}^{+0.043}$ & $3.035$ & $3.043_{-0.036}^{+0.043}$ & $3.072$\\
$n_s$ & $[0.5,1.5]$ & $0.9629_{-0.0059}^{+0.0057}$ & $0.9660$ & $0.9637_{-0.0057}^{+0.0056}$ & $0.9685$\\
$r$ & $[0,1]$ & $0.045_{-0.045}^{+0.011}$ & $0.044$ & $0.057_{-0.057}^{+0.015}$ & $0.061$\\
\hline
$H_0$ & $73.8\pm 2.4$ & $68.08_{-0.74}^{+0.74}$ & $68.28$ & $68.13_{-0.75}^{+0.77}$ & $68.50$\\
$r_{0.002}$ & $...$ & $0.041_{-0.041}^{+0.010}$ & $0.040$ & $0.053_{-0.053}^{+0.014}$ & $0.057$\\
$r_{0.01}$ & $...$ & $0.043_{-0.043}^{+0.010}$ & $0.042$ & $0.055_{-0.055}^{+0.014}$ & $0.059$\\
$\Omega_\Lambda$ & $...$ & $0.6964_{-0.0098}^{+0.0097}$ & $0.7005$ & $0.6971_{-0.0098}^{+0.0099}$ & $0.7022$\\
$\Omega_m$ & $...$ & $0.3036_{-0.0097}^{+0.0097}$ & $0.2995$ & $0.3029_{-0.0099}^{+0.0098}$ & $0.2978$\\
$z_{re}$ & $...$ & $8.8_{-1.5}^{+2.2}$ & $8.8$ & $9.1_{-1.5}^{+2.2}$ & $10.6$\\
${\rm{Age}}/{\rm{Gyr}}$ & $...$ & $13.799_{-0.037}^{+0.036}$ & $13.806$ & $13.798_{-0.038}^{+0.037}$ & $13.785$\\
\hline
$D^{BB}_{l=80}\times 10^{-2}$ & $1.32\pm 0.29/\pm 0.28$ & $0.96_{-0.19}^{+0.19}$ & $0.91$ & $0.80_{-0.23}^{+0.23}$ & $    0.82$\\
$A^{BB}/A^{EE}$ & $0.53\pm0.02$ & $0.53_{-0.02}^{+0.02}$ & $0.53$ & $0.53_{-0.02}^{+0.02}$ & $0.53$\\
$\alpha^{XX}$ & $-2.42\pm0.02$ & $-2.41_{-0.02}^{+0.02}$ & $-2.42$ & $-2.41_{-0.02}^{+0.02}$ & $-2.41$\\
\hline\hline                                                                                                                    
\end{tabular}                                                                                                                   
\caption{The priors, mean values with $1\sigma$ errors and the best fit values of the model parameters and derived cosmological parameters by using {\it Planck}2013+{\it Planck}DUST+WMAP9+BICEP2+BAO+HST data points for $\Lambda$CDM+$r$ model with and without the extrapolation error bars.}\label{tab:results}                                                                                                 
\end{table}                                                                                                                     
\end{center}                                                                                                                    
\endgroup   
\end{widetext}

As shown in Table \ref{tab:results} and Figure \ref{fig:contourI}, by joining BICEP2+{\it Planck}2013+{\it Planck}DUST+WMAP9+BAO+HST, we found the bounds to the tensor-to-scalar ratio $r<0.108$ at $95\%$ confidence level without extrapolating error bars and $r<0.129$ at $95\%$ confidence level with extrapolating error bars for the $\Lambda$CDM+$r$ model. It implies no significant evidence of the primordial gravitational waves. Our results are consistent with that obtained in \cite{ref:Huangdust}. However, as a comparison to the results obtained in \cite{ref:Huangdust}, a relative larger bound was obtained in this work. It is due to the fact that we have considered the propagation of errors for the dust polarization model parameters $D^{BB}_{l=80}$, $A^{BB}/A^{EE}$ and $\alpha^{XX}$. From Table \ref{tab:results} and the Figure \ref{fig:contourI}, one can read off that the tensor-to-scalar ratio $r$ peaks around at a small positive value instead of $0$ as shown in \cite{ref:Huangdust}. And the deviation becomes larger when the extrapolating error bars are include. It might imply a faint signal of the primordial gravitational waves.    

\begin{center}
\begin{figure}[tbh]
\includegraphics[width=9cm]{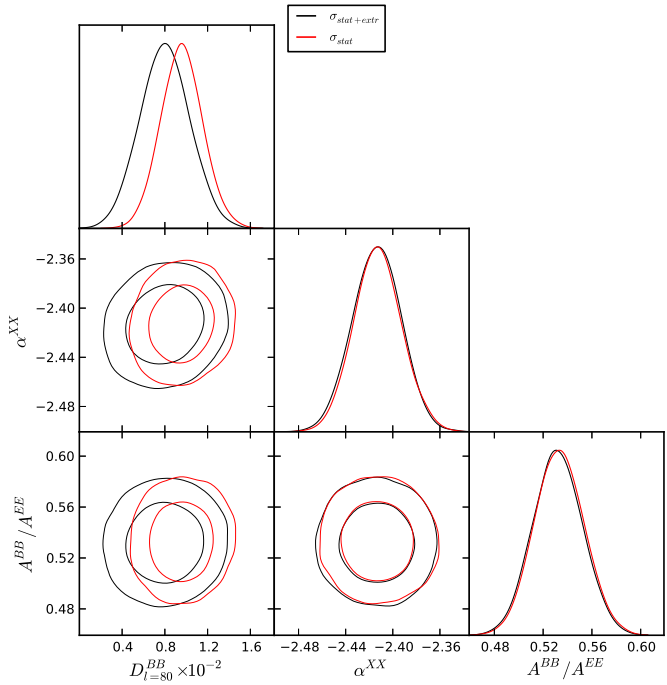}
\caption{The 1D marginalized distribution and 2D contours for HFI $353$GHz dust polarization model parameters by using {\it Planck}2013+{\it Planck}DUST+WMAP9+BICEP2+BAO+HST data points for $\Lambda$CDM+$r$ model with and without the extrapolation error bars.}\label{fig:dustparas}
\end{figure}
\end{center}

In Figure \ref{fig:dustparas}, we show 1D marginalized distribution and 2D contours for HFI $353$GHz dust polarization model parameters obtained from our data analysis. One should notice that the values of $D^{BB}_{l=80}=(0.96\pm 0.19)\times 10^{-2}\mu K^2$ are smaller than that suggested in \cite{ref:Planckdust}. And when the extrapolation error bars are included, the value of $D^{BB}_{l=80}$ is decreed. It is the main reason of the peak of $r$ around a small positive value instead of $0$. It implies the importance to calibrate the amplitude of polarizations of the dust for detecting the primordial gravitational waves in the future.

\begin{center}
\begin{figure}[tbh]
\includegraphics[width=9cm]{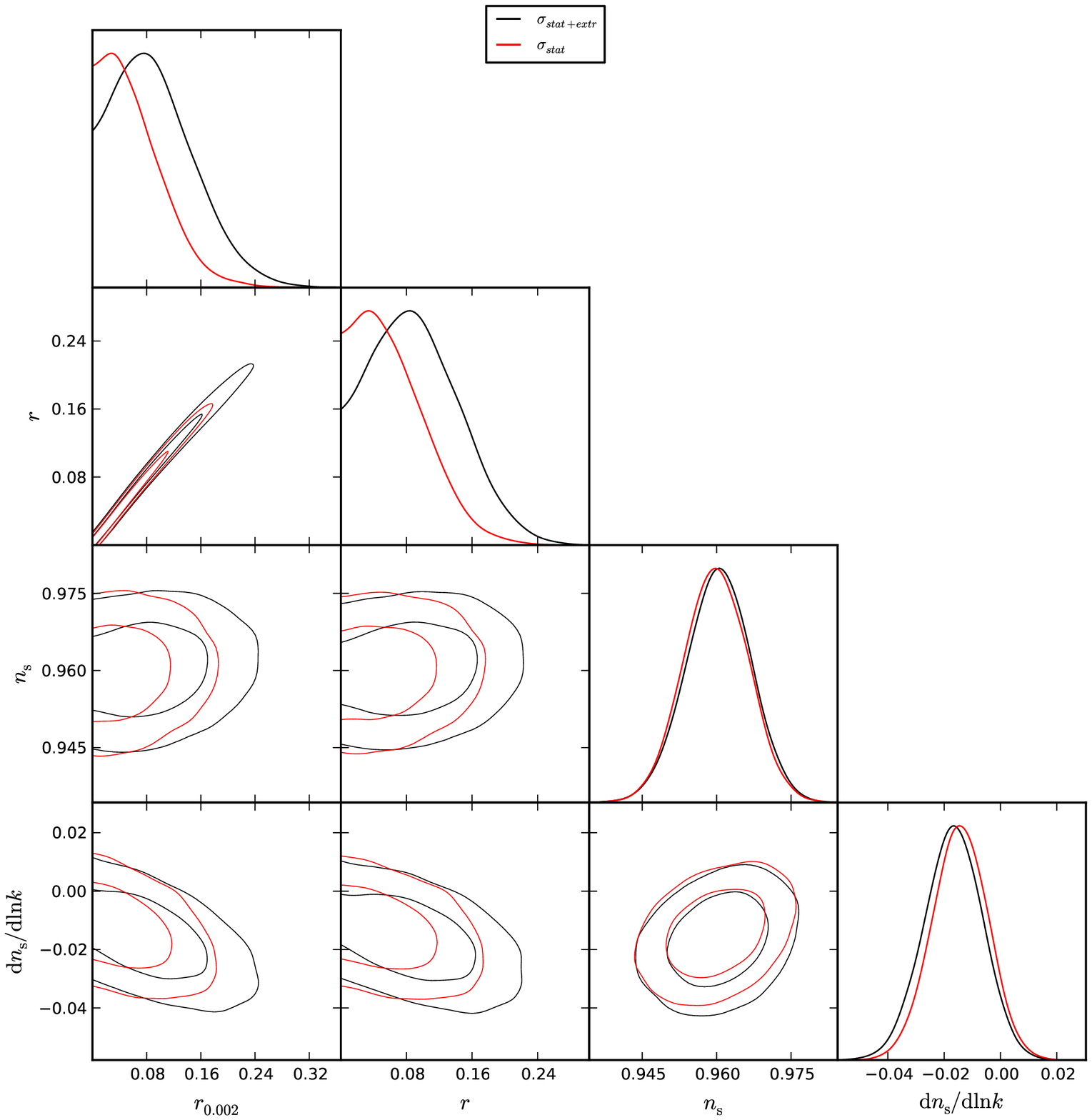}
\caption{The same as Figure \ref{fig:contourI} but including  the running of the scalar spectral tilt.}\label{fig:contournrun}
\end{figure}
\end{center}

\begin{center}
\begin{figure}[tbh]
\includegraphics[width=9cm]{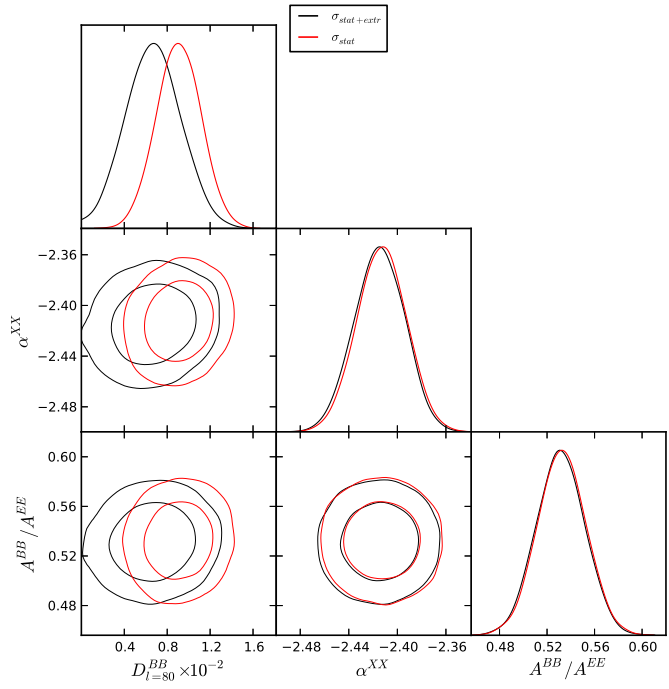}
\caption{The same as Figure \ref{fig:dustparas} but including  the running of the scalar spectral tilt.}\label{fig:dustparasnrun}
\end{figure}
\end{center}

When the running of the scalar spectral tilt is included,  the results are gathered in Table \ref{tab:resultsruning}, Figure \ref{fig:contournrun} and Figure \ref{fig:dustparasnrun} though the same data combination. One can see that the inclusion of the running of the scalar spectral tilt makes the tensor-to-scalar ratio $r$ and reionization redshift $z_{re}$ larger. When the statistical and extrapolating error bars $\sigma_{stat+extr}=0.29+0.28$  are considered, the primordial gravitational waves can be detected in $1\sigma$ regions at the cost of decreasing the value of $D^{BB}_{l=80}$ to $0.67_{-0.25}^{+0.25}$. It again implies the importance to calibrate the amplitude of polarizations of the dust for detecting the primordial gravitational waves.

\begin{widetext}
\begingroup                                                                                                                     
\begin{center}                                                                                                                  
\begin{table}[tbh]                                                                                                                   
\begin{tabular}{cccccc}                                                                                                            
\hline\hline                                                                                                                    
Parameters & Priors & Mean with errors ($\sigma_{stat}$) & Best fit  ($\sigma_{stat}$) & Mean with errors ($\sigma_{stat+extr}$) & Best fit  ($\sigma_{stat+extr}$) \\ \hline
$\Omega_b h^2$ & $[0.005,0.1]$  & $0.02227_{-0.00027}^{+0.00027}$ & $0.02238$ & $0.02230_{-0.00028}^{+0.00028}$ & $0.02227$\\
$\Omega_c h^2$ & $[0.01,0.99]$ & $0.1181_{-0.0017}^{+0.0016}$ & $0.1184$ & $0.1180_{-0.0017}^{+0.0016}$ & $0.1178$\\
$100\theta_{MC}$ & $[0.5,10]$ & $1.04156_{-0.00057}^{+0.00057}$ & $1.04162$ & $1.04159_{-0.00056}^{+0.00056}$ & $1.04165$\\
$\tau$ & $[0.01,0.8]$ & $0.079_{-0.021}^{+0.021}$ & $0.095$ & $0.083_{-0.020}^{+0.021}$ & $0.083$\\
${\rm{ln}}(10^{10} A_s)$ & $[2.7,4.0]$ & $3.066_{-0.042}^{+0.042}$ & $3.102$ & $3.075_{-0.041}^{+0.042}$ & $3.075$\\
$n_s$ & $[0.5,1.5]$ & $0.9599_{-0.0063}^{+0.0064}$ & $0.9563$ & $0.9604_{-0.0064}^{+0.0064}$ & $0.9642$\\
$n_{\rm run}$ & $[-1,1]$ & $-0.0144_{-0.0096}^{+0.0097}$ & $-0.0200$ & $-0.0169_{-0.0101}^{+0.0101}$ & $-0.0129$\\
$r$ & $[0,1]$ & $0.063_{-0.063}^{+0.016}$ & $0.008$ & $0.091_{-0.069}^{+0.042}$ & $0.104$\\
\hline
$H_0$ & $73.8\pm 2.4$ & $68.15_{-0.75}^{+0.74}$ & $68.18$ & $68.22_{-0.75}^{+0.77}$ & $68.29$\\
$r_{0.002}$ & $...$ & $0.062_{-0.062}^{+0.015}$ & $0.008$ & $0.093_{-0.082}^{+0.034}$ & $0.103$\\
$r_{0.01}$ & $...$ & $0.061_{-0.061}^{+0.015}$ & $0.008$ & $0.090_{-0.0721}^{+0.039}$ & $0.102$\\
$\Omega_\Lambda$ & $...$ & $0.6961_{-0.0096}^{+0.0104}$ & $0.6959$ & $0.6969_{-0.0098}^{+0.0099}$ & $0.6983$\\
$\Omega_m$ & $...$ & $0.3039_{-0.01046}^{+0.0096}$ & $0.3041$ & $0.3031_{-0.0099}^{+0.0098}$ & $0.3017$\\
$z_{re}$ & $...$ & $9.9_{-1.7}^{+2.1}$ & $11.4$ & $10.3_{-1.6}^{+2.0}$ & $10.4$\\
${\rm{Age}}/{\rm{Gyr}}$ & $...$ & $13.780_{-0.039}^{+0.039}$ & $13.768$ & $13.776_{-0.039}^{+0.040}$ & $13.777$\\
\hline
$D^{BB}_{l=80}\times 10^{-2}$ & $1.32\pm 0.29/\pm 0.28$ & $0.91_{-0.20}^{+0.20}$ & $1.18$ & $0.67_{-0.25}^{+0.25}$ & $    0.71$\\
$A^{BB}/A^{EE}$ & $0.53\pm0.02$ & $0.53_{-0.02}^{+0.02}$ & $0.54$ & $0.53_{-0.02}^{+0.02}$ & $0.54$\\
$\alpha^{XX}$ & $-2.42\pm0.02$ & $-2.41_{-0.02}^{+0.02}$ & $-2.41$ & $-2.41_{-0.02}^{+0.02}$ & $-2.40$\\
\hline\hline                                                                                                                    
\end{tabular}                                                                                                                   
\caption{The same as Table \ref{tab:results} but including the running of the scalar spectral tilt.}\label{tab:resultsruning}                                                                                                 
\end{table}                                                                                                                     
\end{center}                                                                                                                    
\endgroup   
\end{widetext}

\section{Conclusion} \label{sec:conclusion} 

In this work, we reanalyzed the constraint to the tensor-to-scalar ratio $r$ by joining {\it Planck}2013+{\it Planck}DUST+WMAP9+BICEP2+BAO+HST data points for $\Lambda$CDM+$r$ model. Here in the data analysis, the error bars of the dust polarization parameters $D^{BB}_{l=80}$, $A^{BB}/A^{EE}$ and $\alpha^{XX}$ were included. As a result, we obtained the bound to $r<0.108$ at $95\%$ confidence level without extrapolating error bars and $r<0.129$ at $95\%$ confidence level with extrapolating error bars for the $\Lambda$CDM+$r$ model which are relative larger than that obtained in \cite{ref:Huangdust}. 
As a comparison to the results obtained in \cite{ref:Huangdust}, where $r$ is central at $r=0$, we found a small positive value instead of zero in this work. And the positive value becomes larger when the extrapolation error bars were included. It might imply very weak signal of the primordial gravitational waves. But one should notice that the values of $D^{BB}_{l=80}$ are smaller than that suggested in \cite{ref:Planckdust}. It implies that the calibration of the amplitude of polarizations of the dust is crucial to pin down the detection of the primordial gravitational waves. However when the running of the scalar spectral tilt is included, the values of the tensor-to-scalar ratio $r$ and reionization redshift $z_{re}$ are increased. Even the primordial gravitational waves can be detected in $1\sigma$ regions at the cost of decreasing the value of $D^{BB}_{l=80}$ to $0.67_{-0.25}^{+0.25}$, when statistical and extrapolating error bars $\sigma_{stat+extr}=0.29+0.28$  are considered      

\acknowledgements{When we were preparing the text of this paper, Ref. \cite{ref:Huangdust} appeared on arXiv. But in this work, we considered the error bars of the dust polarization parameters $D^{BB}_{l=80}$, $A^{BB}/A^{EE}$ and $\alpha^{XX}$. As a comparison to their work, we show the very weak signal of the primordial gravitational waves. This work is supported in part by NSFC under the Grants No. 11275035 and "the Fundamental Research Funds for the Central Universities" under the Grants No. DUT13LK01.}

\end{document}